\documentclass{emulateapj}
\usepackage{times}
\usepackage{graphicx}
\usepackage{subfig}
\usepackage{flafter}
\usepackage{natbib}

\begin{document}
\title{\emph{Herschel}\footnotemark[*] Exploitation of Local Galaxy Andromeda (HELGA) III: The Star Formation Law in M31}
\author{George P. Ford\altaffilmark{1}, Walter K. Gear\altaffilmark{1}, Matthew W. L. Smith\altaffilmark{1}, Steve A. Eales\altaffilmark{1}, Maarten Baes\altaffilmark{2}, George J. Bendo\altaffilmark{3}, M\'{e}d\'{e}ric Boquien\altaffilmark{4}, Alessandro Boselli\altaffilmark{4}, Asantha R. Cooray\altaffilmark{5}, Ilse De Looze\altaffilmark{2}, Jacopo Fritz\altaffilmark{2}, Gianfranco Gentile\altaffilmark{2,6}, Haley L. Gomez\altaffilmark{1}, Karl D. Gordon\altaffilmark{2,7}, Jason Kirk\altaffilmark{1}, Vianney Lebouteiller\altaffilmark{8}, Brian O'Halloran \altaffilmark{9}, Luigi Spinoglio\altaffilmark{10}, Joris Verstappen\altaffilmark{2}, Christine D. Wilson\altaffilmark{11}}

\shorttitle{HELGA III: The Star Formation Law in M31}

\altaffiltext{1}{School of Physics and Astronomy, Cardiff University, Queens Buildings, The Parade, Cardiff, CF24 3AA, United Kingdom}
\altaffiltext{2}{Sterrenkundig Observatorium, Universiteit Gent, Krijgslaan 281, B-9000 Gent, Belgium}
\altaffiltext{3}{Jodrell Bank Centre for Astrophysics, University of Manchester, Alan Turing Building, Manchester, M13 9PL, United Kingdom}
\altaffiltext{4}{Aix Marseille Universit\'{e}, CNRS, LAM (Laboratoire d'Astrophysique de Marseille) UMR 7326, 13388, Marseille, France}
\altaffiltext{5}{University of California, Irvine, Department of Physics \& Astronomy, 4129 Frederick Reines Hall, Irvine, CA 92697-4575, United States of America}
\altaffiltext{6}{Department of Physics and Astrophysics, Vrije Universiteit Brussel, Pleinlaan 2, 1050 Brussels, Belgium}
\altaffiltext{7}{Space Telescope Science Institute, 3700 San Martin Drive, Baltimore, MD 21218, USA}
\altaffiltext{8}{Service d’Astrophysique, l’Orme des Merisiers, CEA, Saclay, France}
\altaffiltext{9}{Imperial College London, Astrophysics, Blackett Laboratory, Prince Consort Road, London SW7 2AZ, UK}
\altaffiltext{10}{INAF, Istituto di Fisica dello Spazio Interplanetario, Via Fosso del Cavaliere 100, Tor Vergata, IT 00133 Roma, Italy}
\altaffiltext{11}{Department of Physics \& Astronomy, ABB-241, McMaster University, 1280 Main St. W, Hamilton, ON, L8S 4M1, Canada}

\begin{abstract}
We present a detailed study of how the Star Formation Rate (SFR) relates to the interstellar medium (ISM) of M31 at $\sim140\rm\,pc$ scales. The SFR is calculated using the far-ultraviolet and 24\,$\rm \mu m\,$ emission, corrected for the old stellar population in M31. We find a global value for the SFR of $0.25^{+0.06}_{-0.04}\,\rm M_{\odot}\,yr^{-1}\,$ and compare this with the SFR found using the total far-infrared (FIR) luminosity. There is general agreement in regions where young stars dominate the dust heating. Atomic hydrogen (H\,{\sc i}) and molecular gas (traced by carbon monoxide, CO) or the dust mass is used to trace the total gas in the ISM. We show that the global surface densities of SFR and gas mass place M31 amongst a set of low-SFR galaxies in the plot of \citet{kennicutt1998b}. The relationship between SFR and gas surface density is tested in six radial annuli across M31, assuming a power law relationship with index, $N$. The star formation law using total gas traced by H\,{\sc i} and CO gives a global index of $N=2.03\pm0.04$, with a significant variation with radius; the highest values are observed in the $10\,{\rm kpc}$ ring. We suggest that this slope is due to H\,{\sc i} turning molecular at $\Sigma_{\rm Gas}\sim10\,\rm M_{\odot}\,pc^{-2}$. When looking at H$_{2}$ regions, we measure a higher mean SFR suggesting a better spatial correlation between H$_{2}$ and SF. We find $N\sim0.6$ with consistent results throughout the disk - this is at the low end of values found in previous work and argues against a superlinear SF law on small scales.
\end{abstract}

\section{Introduction}
Relationships between star formation and the quantity of gas in the interstellar medium (ISM) seem to vary greatly across the universe \citep{kennicutt2012}. There is a tight correlation between Star Formation Rate (SFR) and gas mass in most cases, but the slope of the relationship does not appear to be consistent. A specified surface density of gas can correspond to a range of SFR even in a single galaxy.

Schmidt's original paper \citep{schmidt1959}, which focussed on the Milky Way, derived a relationship between the volume density of star formation  and the volume density of gas ($\rho_{\rm SFR} \sim \rho_{\rm Gas}^N$) with a power law index $N = 2$. The first extragalactic measurements of the Schmidt law were carried out by \citet{sanduleak1969} and \citet{hartwick1971} on the Small Magellanic Cloud (SMC) and M31 respectively and found indices of $N_{\rm SMC} = 1.84 \pm 0.14$ and $N_{\rm M31} = 3.50 \pm 0.12$.

Since then, similar studies have tended to relate surface density of star formation ($\Sigma_{\rm SFR}$) to surface density of gas ($\Sigma_{\rm Gas}$), which are what we actually observe. However, the index probed using surface densities should be equivalent to that for volume densities as long as we have a constant scale height.

A later study of 16 nearby galaxies by \citet{boissier2003} found $N = 2$. \citet{wong2002} studied 6 nearby spirals and estimated $N$ to be in the range 1.2-2.1. However \citet{heyer2004} calculated an index of $\sim 3.3$ for M33 when considering total gas, but $N \sim 1.4$ when looking only at molecular hydrogen. A more recent work on the same object \citep{verley2010} find a wide range of indices ($1.0<N<2.6$) depending on gas tracer and fitting method. In the comprehensive (and most often cited) work of \citet{kennicutt1998b}, he estimated the power index for 90 nearby galaxies using total gas (molecular gas only for starbursts) and found $N = 1.40 \pm 0.15$. A big question is whether this slope only works when considering global measurements or is it a manifestation of a relationship on smaller scales.

One interpretation of the Kennicutt result is that star formation timescales are dictated by the free-fall time, \hbox{${\rm SFR} \sim M / \tau_{\rm ff} \sim \rho / \rho^{-1/2} \sim \rho^{3/2}$}, since $\tau_{\rm ff} \propto \rho^{-1/2}$ \citep[e.g.][]{elmegreen1994, krumholz2007, narayanan2008}. Other work suggests that the super-{\footnotesize \\ \\ $^{*}${\textit{H\lowercase{erschel}} \lowercase{is an} ESA \lowercase{space observatory with science instruments provided by} E\lowercase{uropean-led} P\lowercase{rincipal} I\lowercase{nvestigator consortia and with important participation from} NASA.}} \\linear slope is a result of variations in the fraction of dense gas between normal spiral galaxies and starbursts and that the star formation law is linear given constant dense gas fraction \citep[e.g.][]{lada2012}.

Recently acquired data from the Galaxy Evolution Explorer (GALEX, \citealt{galex}) and The H\,{\sc i} Nearby Galaxies Survey (THINGS) has allowed the star formation law to be probed on sub-kpc ($\sim750\,$pc) scales as in \citet{bigiel2008}. They suggest that star formation is more directly related to molecular rather than total gas and find that the molecular gas star formation law follows a relationship with index, $N = 1$ (a linear relationship), consistently lower than the values they find for total gas.

M31, due to its proximity and size ($D = 785{\,\rm kpc}$ \citep{mcconnachie2005}, apparent angular size is 190\,\arcmin) gives the opportunity to probe these relationships over an entire galaxy, at smaller physical scales (comparable to the size of a giant molecular cloud) than all other extragalactic spirals. It is also roughly solar metallicity \citep{yin2009} so is a good analogue to the Milky Way. This has been done previously (on lower fidelity data than here) by \citet{tabatabaei2010}, who found a similar super-linear relationship between surface densities of star formation from H$\alpha\,$ and total gas to that found in \citet{kennicutt1998b} for whole galaxies.

Star formation tracers, whether looking at unobscured or embedded star formation, invariably rely on the assumption that the emission used as a SF probe originates directly, or as a result of heating, from young stars \citep{calzetti2007b}. This is a reasonable assumption in galaxies that have recently undergone a starburst, as massive young stars burn brightly and die young, with less massive stars living much longer and providing a minimal contribution to the ultraviolet luminosity. However, M31 has not undergone a recent starburst \citep{davidge2012} so contributions from older populations can have a significant effect on star formation estimates \citep{calzetti2012}. This should, in principle, be possible to mitigate using tracers of the general stellar population.

In this paper, we present multi-wavelength data of M31 and measure the total unobscured and embedded star formation rates separately using far-ultraviolet (FUV) and 24\,\micron\, data respectively. We compare this with the total gas, found by combining maps of neutral atomic hydrogen (H\,{\sc i}) and carbon monoxide (CO($J$=1-0)), which traces the molecular hydrogen (H$_{2}$).

The maps tracing SFR and gas mass are used to calibrate the SFR and gas mass found using the far infrared (FIR) emission from M31, as observed with the \emph{Herschel  Space Observatory} \citep{pilbratt2010} as part of the \emph{Herschel} Exploitation of Local Galaxy Andromeda (HELGA) project (\citealt{helgai}). We compare our SFR from UV and 24\,\micron\, emission with that found from FIR luminosity. The interstellar gas mass is also traced using the dust mass estimated from the FIR Spectral Energy Distribution (SED), scaled using the observed gas to dust ratio. Here we aim to see how well this gas map correlates with SFR, hence whether dust mass traces star forming regions. 

Finally, we use this collection of SFR and gas maps to probe the power law relationship between SFR surface density and the gas surface density, or Kennicutt-Schmidt (K-S) law. Our analysis is performed on individual pixels in M31 and investigates how the law varies with different gas tracers on sub-kpc scales.

\section{Data}
\begin{figure*}[!htb]
  \centering
    \includegraphics[width=0.55\textwidth, trim=0 190 1136.25 2036.475, clip=true]{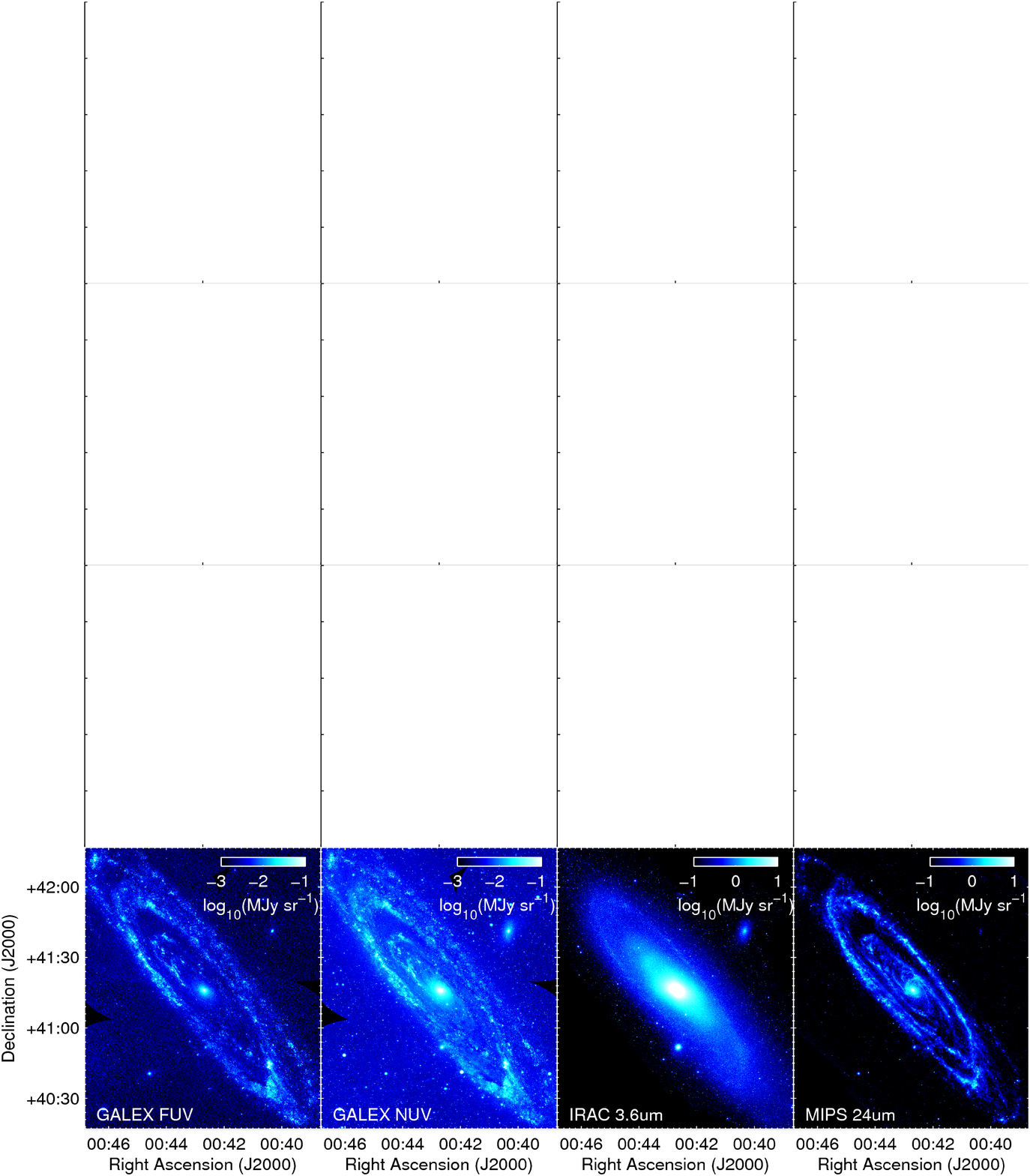}\includegraphics[width=0.225\textwidth, trim=1388.75 190 568.125 2036.475, clip=true]{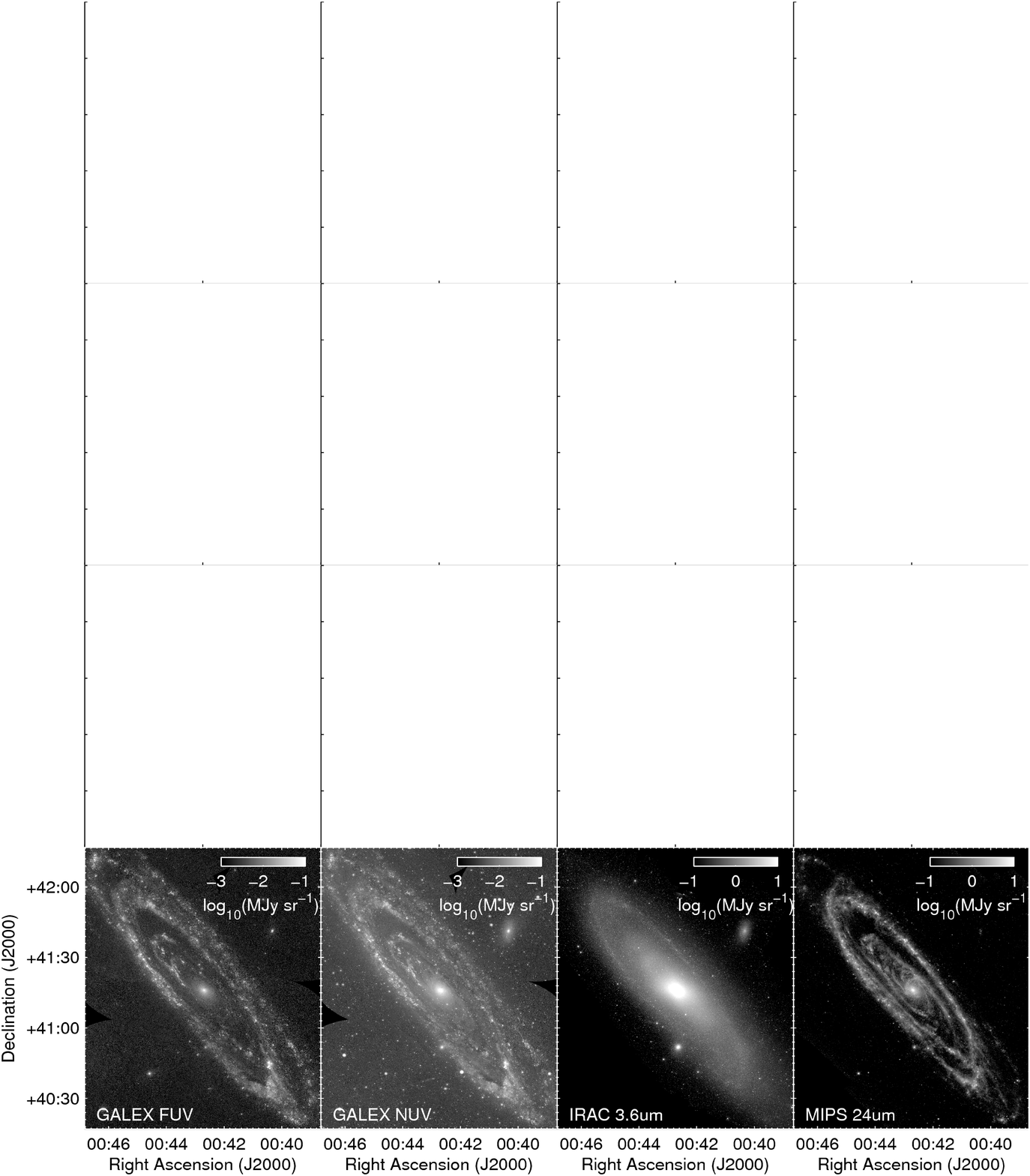}\includegraphics[width=0.225\textwidth, trim=1956.875 190 0 2036.475, clip=true]{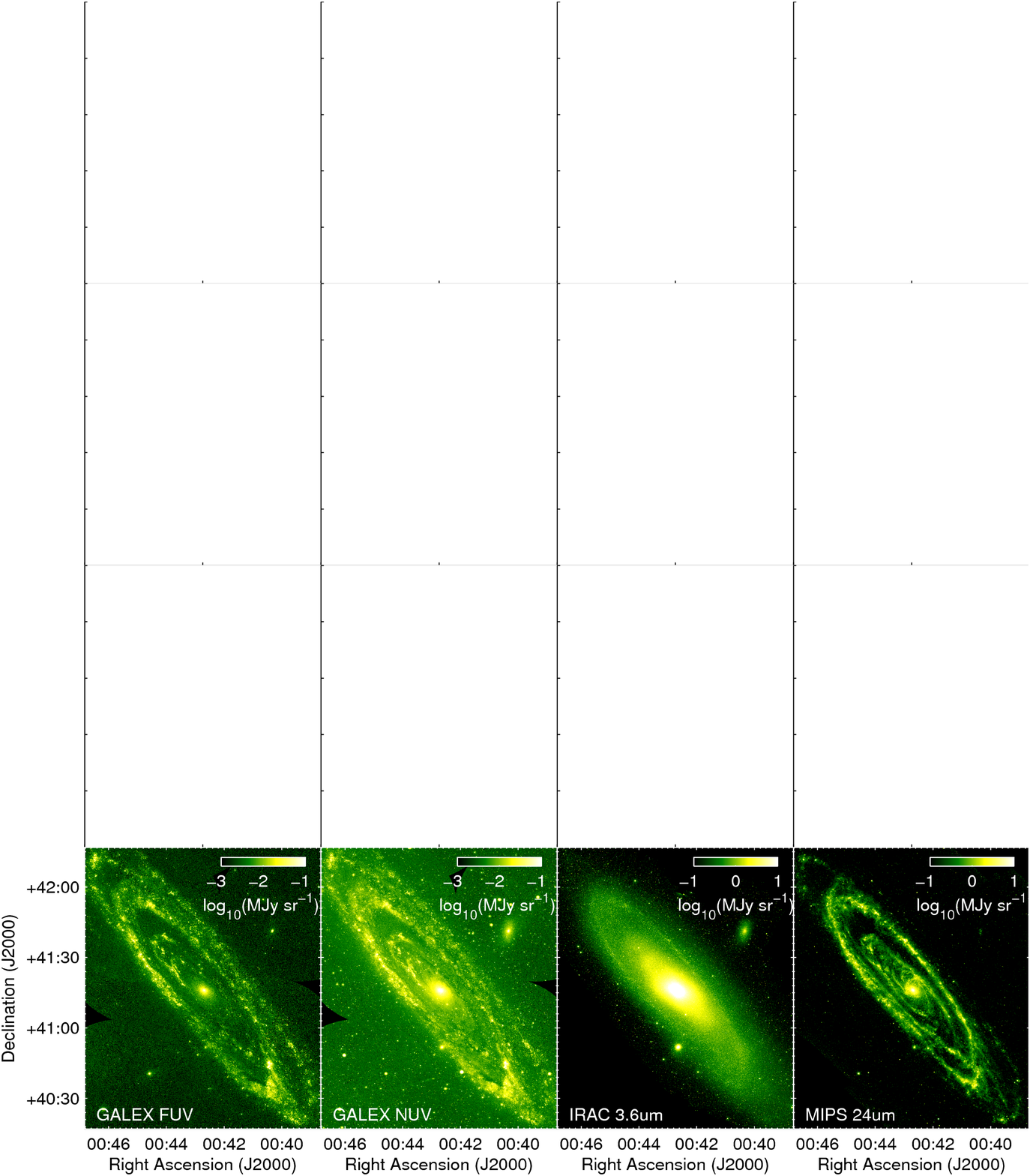}
  \caption{Images used in the creation of the FUV and 24\micron\, star formation map of M31. From left: \emph{GALEX} FUV and NUV maps \citep{galex}, \emph{Spitzer IRAC} 3.6\,\micron\, \citep{barmby2006}, \emph{Spitzer MIPS} 24\,\micron\, \citep{gordon2006}.}
\label{fig:mapsL08}
\end{figure*}

Our first method of tracing star formation uses \emph{GALEX} FUV and NUV observations of M31 \citep{galex}, along with warm dust emission seen in \emph{Spitzer MIPS} 24\,\micron\, \citep{gordon2006} and stellar emission from \emph{Spitzer IRAC} 3.6\,\micron\, \citep{barmby2006} (Figure \ref{fig:mapsL08}).

% FIGURE 1

\begin{figure*}[!htb]
  \centering
    \includegraphics[width=\textwidth, trim=0 133.333 0 612, clip=true]{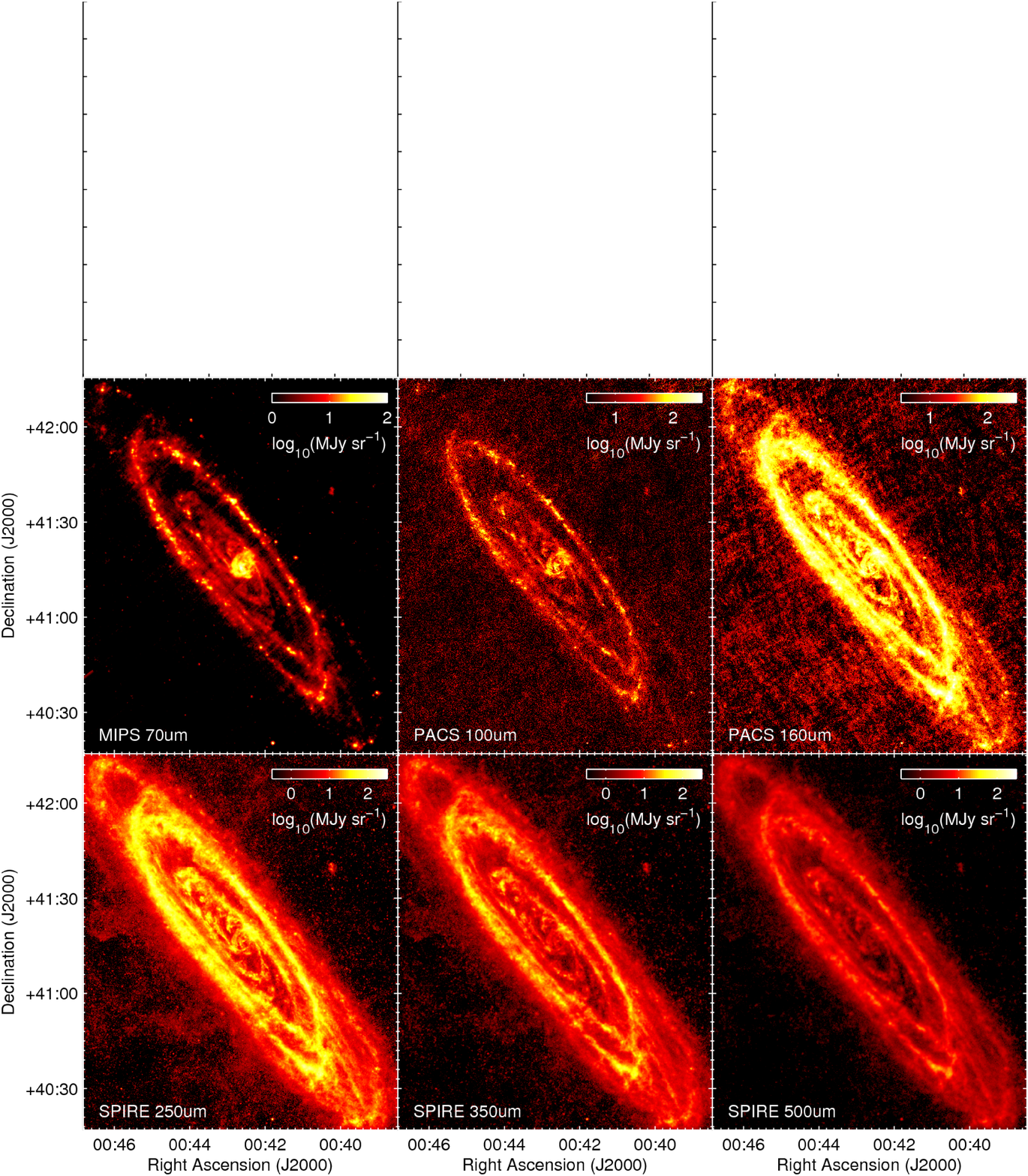}
  \caption{Top, from left: \emph{Spitzer MIPS} 70\,\micron\, \citep{gordon2006}, \emph{PACS} 100 and 160\,\micron\, \citep{helgai}; bottom: \emph{SPIRE} 250, 350 and 500\,\micron\, \citep{helgai}. The \emph{PACS} and \emph{SPIRE} observations are from the HELGA collaboration.}
\label{fig:mapsIR}
\end{figure*}

The HELGA collaboration obtained observations of M31 in five \emph{Herschel} bands \citep{helgai}. They are \emph{PACS} \citep{poglitsch2010} 100 and 160\,\micron\, and \emph{SPIRE} \citep{griffin2010} 250, 350 and 500\,\micron. Details of the data reduction for both \emph{PACS} and \emph{SPIRE} maps can be found in \citet{helgai}. The \emph{Spitzer MIPS} 70\,\micron\, map \citep{gordon2006}, is employed to extend the wavelength range for our calculation of the FIR spectral energy distribution (Figure \ref{fig:mapsIR}).

% FIGURE 2

\begin{figure*}[!htb]
  \centering
    \includegraphics[width=0.7\textwidth, trim=0 200 772.5 1846, clip=true]{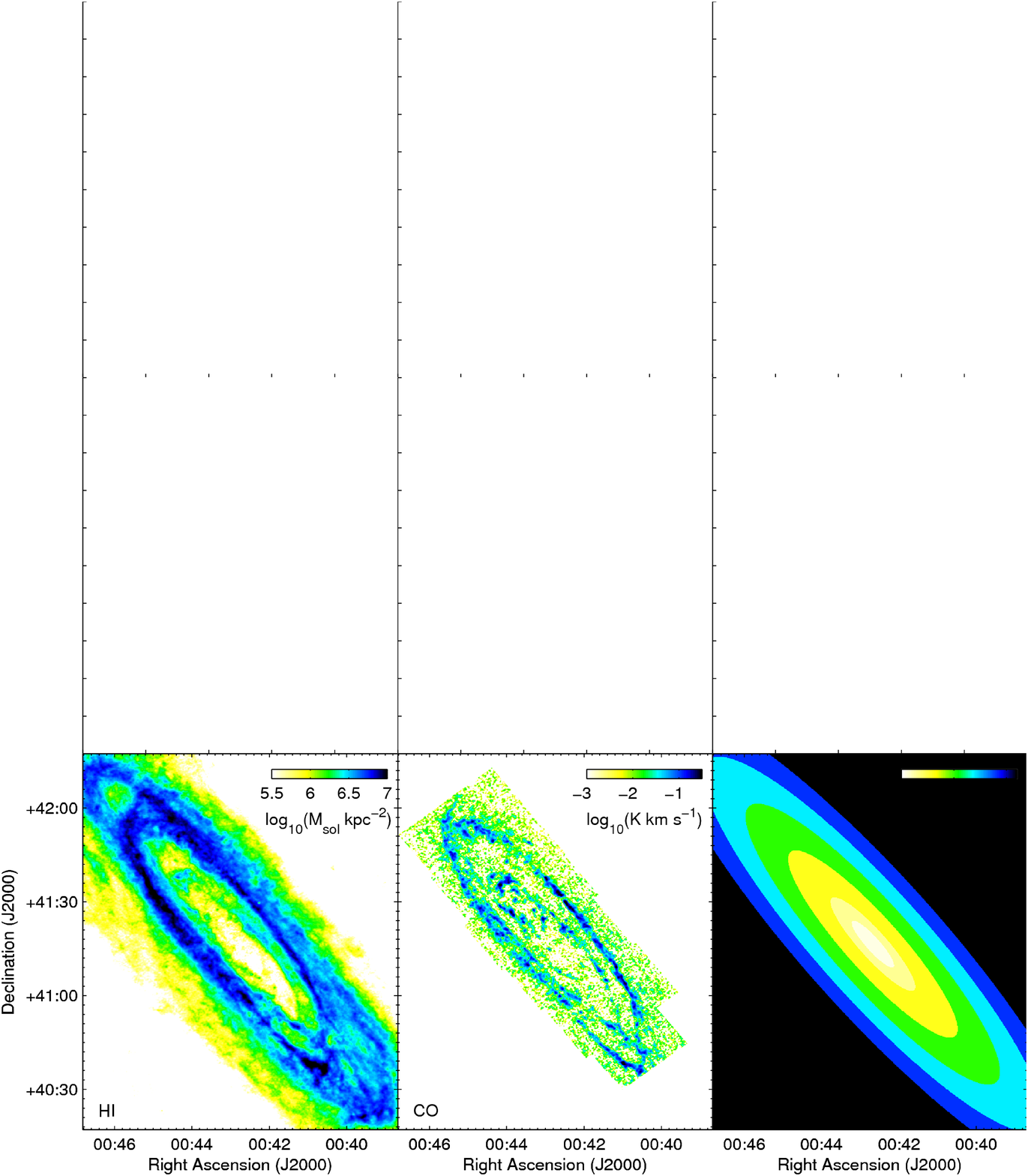}\includegraphics[width=0.3\textwidth, trim=1802.5 200 0 1846, clip=true]{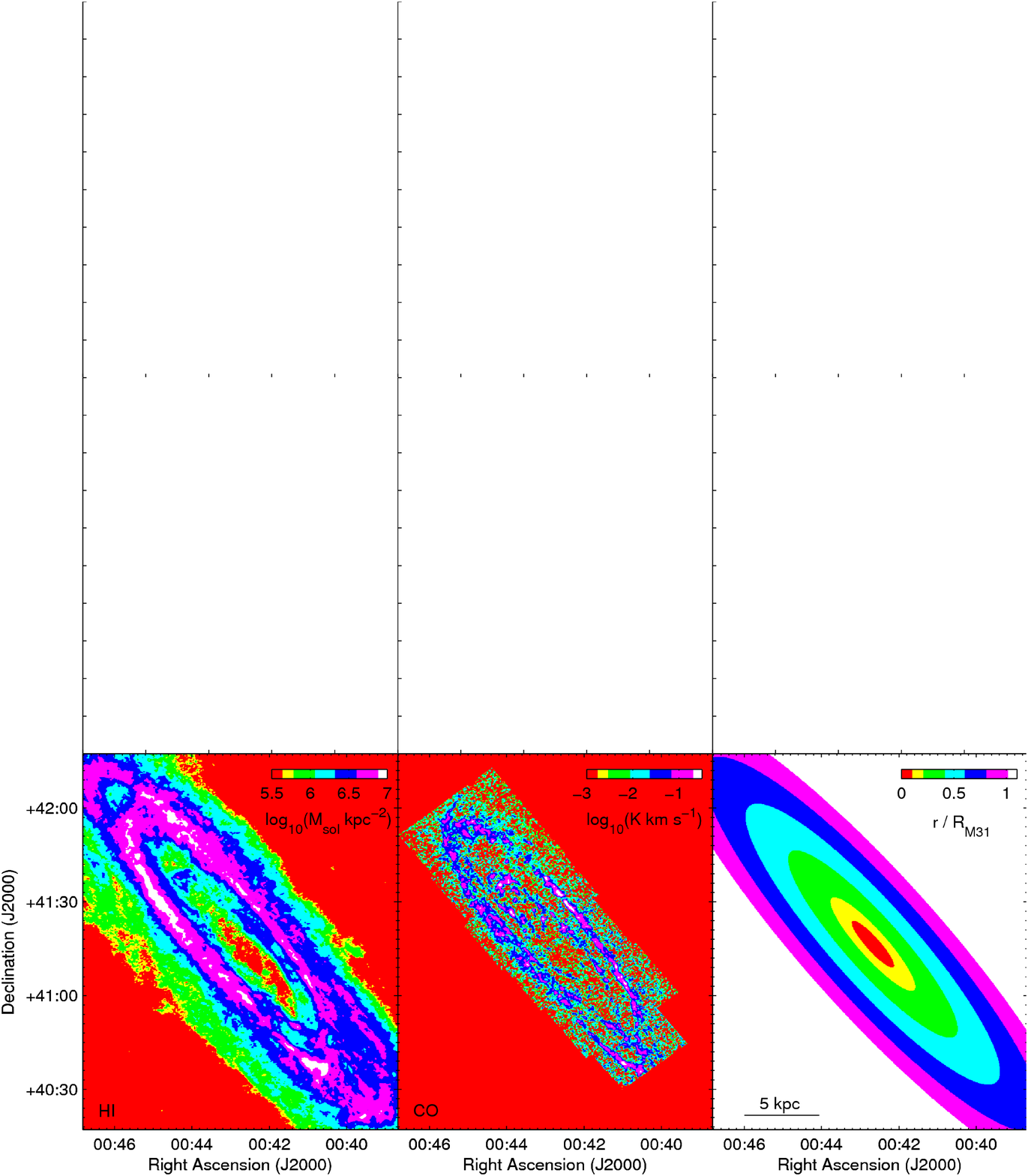}
  \caption{Left, integrated H\,{\sc i} emission \citep{braun2009}; centre, CO($J$=1-0) \citep{nieten2006}; right, colour key of elliptical annuli. Each annulus is 0.2$\,R_{\rm M31}$ thick apart from the inner two which are 0.1$\,R_{\rm M31}$ thick.}
\label{fig:mapsgas}
\end{figure*}

We independently probe the interstellar medium using H\,{\sc i} \citep{braun2009} and CO($J$=1-0) maps \citep{nieten2006} (Figure \ref{fig:mapsgas}). Note that the CO map covers a smaller area than the H\,{\sc i}. The area not covered by the CO will be used in the calculation of total gas, providing there is sufficient H\,{\sc i}.

% FIGURE 3

In this work, we divide the maps into elliptical annuli of constant galactocentric radius. We do this to test the effect of radius on the star formation law, with the option to relate this to the Toomre Q criterion, which relates to rotational velocity and shear. It also allows us to isolate the 10 kpc ring, where the majority of star formation in M31 is occurring; and the central regions which are dominated by an older stellar population. The right-hand image in Figure \ref{fig:mapsgas} shows the colour coding of datapoints used in all subsequent plots depending on their radial distance from the centre. The ellipses are created assuming a position angle of Andromeda of 38$^{\circ}$\, and an inclination of 77$^{\circ}$\, \citep{mcconnachie2005}. The colours indicate how the datapoints from those regions will be represented in later figures. Distances are in units of $R_{\rm M31}$ which we take to be 21.55\,kpc \citep{devaucouleurs1991}.

For analysis, the maps are individually smoothed and regridded to three pixel scales, based on the lowest resolution map used in the analysis. We modify the full-width half-maximum ({\sc FWHM}) beamwidth to match the effective Point Spread Function (PSF) by Gaussian smoothing the image using the \emph{IRAF} function \emph{imgauss}. The maps are regridded using the IDL \emph{astrolib} function, \emph{FREBIN}. Any offsets in the coordinates of the pixels are corrected using \emph{wcsmap} and \emph{geotran} in \emph{IRAF}.

The first scale used here is the highest resolution star formation map we can create using the FUV and 24\,\micron\, emission as a tracer. This corresponds to the lowest resolution (\emph{MIPS} 24\,\micron) {\sc FWHM} beamwidth of 6\,\arcsec\, ($\sigma_{\rm beam} = 2.55\,\arcsec$) and a pixel size of 1.5\,\arcsec. This scale is applied to the 3.6\,\micron, 24\,\micron, NUV and FUV maps.

We aim to study the relationship between SFR and gas mass on the smallest scales attainable. To this end, we also use maps smoothed to the resolution and grid size of the neutral atomic hydrogen map, again the lowest resolution map used here. The effective {\sc FWHM} beamwidth is 30\,\arcsec\, ($\sigma_{\rm beam} = 12.7\,\arcsec$) with a 10\,\arcsec\, pixel size. This scale is applied to the data mentioned above, with the addition of the CO($J$=1-0) map.

In order to compare gas mass (from H\,{\sc i} and CO($J$=1-0)) and star formation in M31 (from FUV and 24\,\micron\, emission) with the \emph{Herschel} observations, the majority of the analysis is performed on a scale corresponding to the beamsize of the lowest resolution SPIRE map (500\,\micron). These images have an effective {\sc FWHM} beamwidth and grid size of 36\,\arcsec\, ($\sigma_{\rm beam} = 15.5\,\arcsec$). Since the beamwidth and pixel size are equivalent, the pixels can be described as approximately `independent,' as there is no correlation between them. Here, the \emph{MIPS} maps (Figure \ref{fig:mapsIR}) were smoothed using convolution kernels from \citet{bendo2012} as described in \citet{smith2012b}.

\newpage
\newpage
\section{Star formation rate}
\label{sec:sfr_create}
\subsection{FUV and 24\,\micron\,}
\label{sec:fuv24}
The star formation rate is first calculated from the \emph{GALEX} FUV and \emph{Spitzer} 24\,\micron\, maps, using the method prescribed in \citet{leroy2008}. However, to expand on this we also use \emph{GALEX} NUV and \emph{Spitzer IRAC} 3.6\,\micron\, maps to correct for foreground stars and emission from old stellar populations respectively.

FUV emission is predominantly from unobscured high-mass stars (O, B and A-type), so this tracer is sensitive to star formation on a timescale of $\sim100$\,Myr \citep[e.g.][]{kennicutt1998a, calzetti2005, salim2007}. 24\,\micron\, emission is predominantly due to dust-heating by UV photons from bright young stars, and is sensitive to a star formation timescale of $<\,$10\,Myr \citep[e.g.][]{calzetti2005, perez-gonzalez2006, calzetti2007}.

The star formation surface density is calculated using the formulation in \citet{leroy2008} which uses a Chabrier initial mass function (IMF):
\begin{equation}
\label{eq:fuv24tosfr}
\Sigma_{\rm SFR} = 8.1\times10^{-2}\,I_{\rm FUV}+3.2^{+1.2}_{-0.7}\times10^{-3}\,I_{24},
\end{equation}
where $\Sigma_{\rm SFR}$ has units of M$_{\odot}\,\rm yr^{-1}\,\rm kpc^{-2}$ and FUV and 24\,\micron\, intensity ($I$) are in $\rm MJy\,sr^{-1}$. The pixel size corresponds to a distance of $\sim140\,pc$. If comparing like for like with other galaxies, an inclination correction factor of $\cos{i}$ (where the inclination of M31, $i = 77^{\circ}$)  must be included in order to `deproject' the image, effectively giving values as they would be for a face-on galaxy. This prescription assumes all the 24\,\micron\, emission in M31 is due to dust heating by newly formed stars, and that the FUV is emitted exclusively by young stars. There are, of course, other sources of these tracers which are unrelated to star formation which must be taken into account.

The first issue is foreground stars. These are selected and removed using the UV colour, as in \citet{leroy2008} --- if $I_{\rm NUV}/I_{\rm FUV}\,>\,15$ the pixel is blanked in both the FUV and 24\,\micron\, map (some 24\,\micron\, emission will be stellar, e.g. \citealt{bendo2006}). We assume this ratio will only be reached where a pixel is dominated by a single star, which, given our resolution will never be associated with M31.

\begin{figure*}[!htb]
  \centering
    \includegraphics[width=\textwidth, trim=5 0 0 5, clip=true]{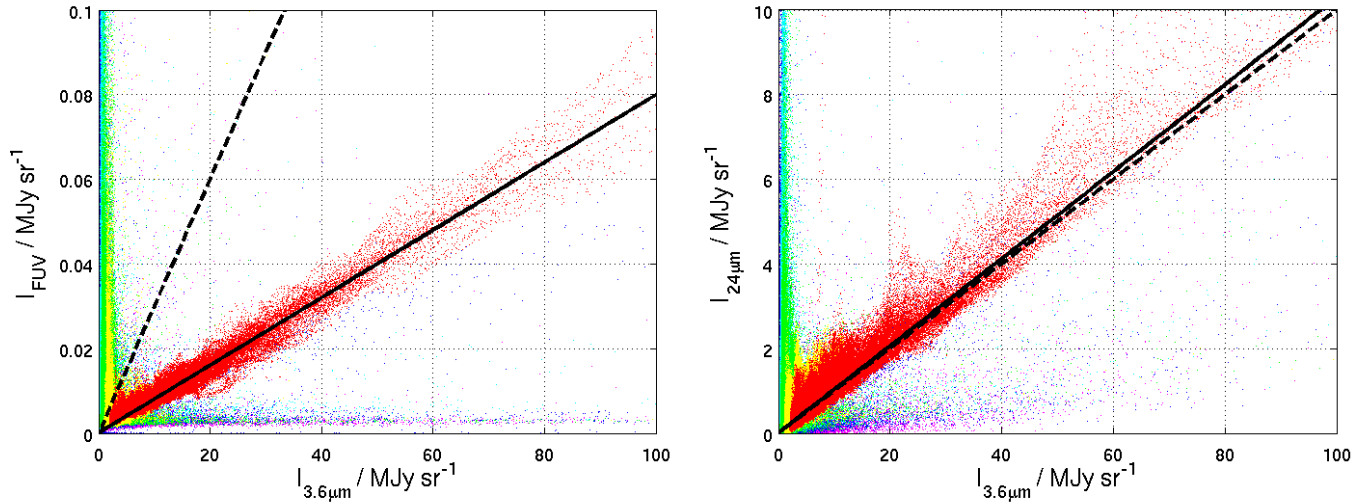}
  \caption[24\,\micron\, and FUV vs 3.6\,\micron\, emission]{In order to study the effect of old stars on our star formation tracers, we plot FUV vs 3.6\,\micron\, and 24\,\micron\, vs 3.6\,\micron\, emission, in this case on the high resolution map. Colours indicate the distance, from the galactic centre, of each pixel (see Figure \ref{fig:mapsgas}). The black dashed trendline indicates the correction for the old stellar population used in \citet{leroy2008}, based on $I_{\rm FUV}/I_{\rm 3.6}$ and $I_{24}/I_{3.6}$ found in ellipticals. The solid trendline is the best fit to FUV vs 3.6\,\micron\, (left) and 24\,\micron\, vs 3.6\,\micron\, (right) in the inner regions of M31 ($r<0.1R_{\rm M31}$). We can see that there is a tight correlation in both plots indicating that FUV and 24\,\micron\, emission in the centre of M31 is predominantly from old stars.}
\label{fig:old_pop}
\end{figure*}

A second problem is that some of the emission could be from an older stellar population. This is a general problem and not specific to M31 \citep[e.g.][]{kennicutt2009}. We expect this to be a bigger issue near the centre of the galaxy. Previous FIR work on M31 \citep{tabatabaei2010} avoids this problem by measuring the SFR at radii greater than 6 kpc only, based on the assumption that the centre of the galaxy contains negligible star formation. Old stars are fainter but redder, so emit relatively stronger at 3.6\,\micron. This means we can mitigate for the old stars by determining $I_{\rm FUV}/I_{\rm 3.6}$ (hereafter, $\alpha_{FUV}$) and $I_{24}/I_{3.6}$ (hereafter, $\alpha_{24}$) in regions where we assume star formation has ceased, and use this to remove the component of FUV and 24\,\micron\, emission coming from old stars. So, the emission we associate with star formation is given by,
\begin{equation}
I_{\rm FUV, SF} = I_{\rm FUV} - \alpha_{\rm FUV}\,I_{3.6}
\end{equation}
\begin{equation}
I_{\rm 24, SF} = I_{\rm 24} - \alpha_{24}\,I_{3.6}.
\end{equation}
\citet{leroy2008} explored this by looking at the ratio of fluxes determined in elliptical galaxies. They found $\alpha_{\rm FUV} = 3 \times 10^{-3}$ and $\alpha_{\rm 24} = 0.1$. However, if we compare the 3.6\,\micron\, emission with the FUV and 24\,\micron\, in M31, we see that these values are not necessarily appropriate here (Figure \ref{fig:old_pop}). The 24\,\micron\, emission in the bulge (shown by red points) follows the ratio found in ellipticals (black-dashed line, Figure \ref{fig:old_pop}, right), so we will use the same value for $\alpha_{\rm 24}$. $\alpha_{\rm FUV}$ is found to be much lower here (Figure \ref{fig:old_pop}, top). We speculate this is due to dust extinction in M31, which is not an issue in passive elliptical galaxies as they contain little dust \citep[e.g.][]{smith2012a, rowlands2012}. It is also stated in \citet{leroy2008} that there is a large scatter in this ratio so a discrepancy is not surprising. An independent correction is found by performing linear fits on the inner regions of M31. Ellipses within a radius 0.05, 0.1 and 0.2$\,R_{\rm M31}$ give gradients ($\alpha_{\rm FUV}$) of $8.42 \times 10^{-4}$, $7.99 \times 10^{-4}$ and $7.44 \times 10^{-4}$ respectively. Here, the mean value, $\alpha_{\rm FUV} = 8.0 \times 10^{-4}$, will be employed to correct FUV emission for the old stellar population in M31. We performed this analysis on the high resolution maps to maximise the number of datapoints and checked that the slope was consistent with that found using the lowest resolution ($36\,\arcsec$) maps.

% FIGURE 4

Once this correction is applied, we have a map of surface density of star formation, $\Sigma_{\rm SFR}$ in units of $\rm M_{\odot}\,yr^{-1}\,kpc^{-2}$ (Figure \ref{fig:mapssfr}). The correction in this work has the effect of reducing the measured global SFR by $\sim25\%$.

It should be noted that when looking at the region immediately outside the 10\,kpc ring only (dark blue points, behind red points (not visible), Figure \ref{fig:old_pop}, left), we see a tight correlation between the FUV and 3.6\,\micron\, emission. Unlike the centre, however, there are pixels in that region that do not follow this correlation. This indicates that despite a significant population of old stars in the ring, star formation is still occurring at significant rates compared to the rest of the galaxy ($\sim0.2\,\rm M_{\odot}\,\rm yr^{-1}$).

We can scale the star formation map to give the total star formation occurring within a pixel. This allows us to determine the global SFR for M31, which we find to be $0.25^{+0.06}_{-0.04}\,\rm M_{\odot}\,yr^{-1}$, almost an order of magnitude lower than the canonical value of $\sim2\,\rm M_{\odot}\,\rm yr^{-1}$, quoted in \citet{chomiuk2011} for the Milky Way but consistent with the lower limit of $\sim0.27\,\rm M_{\odot}\,\rm yr^{-1}$ found in \citet{tabatabaei2010} for M31.

\subsection{Star formation from far-infrared luminosity}
\label{sec:sfr_ir}
Star formation can also be calculated using the total FIR luminosity. This ideally probes the embedded SFR and is sensitive to cooler dust temperatures. This can be an issue in determining total SFR in dust-deficient galaxies where significant starlight is not attenuated by dust but should not be a problem here. An issue that is relevant to M31 is that a significant component of the dust heating could be from an evolved stellar population \citep[e.g.][]{bendo2010, boquien2011, bendo2012, smith2012b}.

The total FIR flux was integrated in frequency space using a linear interpolation between the six datapoints (70-500\,\micron) for each pixel independently. Each value was converted from a flux to a luminosity in $\rm L_{\odot}$ assuming a distance to M31 of 785\,kpc \citep{mcconnachie2005}.

If we assume FIR luminosity is exclusively re-radiated emission from warm dust that is heated during a continuous starburst, the FIR luminosity is equal to the total luminosity of the starburst. The total SFR, $\dot M_{\rm SF}$ is then,
\begin{equation}
\dot M_{\rm SF} = \delta_{\rm MF}~
        (L_{\rm FIR}/\rm 10^{10}L_\odot)~~M_\odot yr^{-1}~,
\label{eq:sfr_ir}
\end{equation}
where $\delta_{\rm MF}$ depends on the assumed IMF of the region being studied and the timescale of the starburst. Changing these assumptions gives radically different conversion factors. For example, assuming a Kroupa IMF with a star formation timescale of $10\,\rm Gyr$ gives $\delta_{\rm MF} \sim 0.6$ whereas the same IMF with a timescale of $2\,\rm Myr$ gives $\delta_{\rm MF} \sim 3.2$ \citep{calzetti2012}.

Here we employ the value quoted in \citet{kennicutt1998b} of $\delta_{\rm MF} = 1.7$, which assumes a Salpeter IMF with a low mass cut-off of $0.1\,\rm M_{\odot}$ and a timescale of $\sim100\rm Myr$. We should state here that this assumes a continuous starburst which keeps consistency with the previous method. This conversion factor gives a global star formation rate of $0.52\,\rm M_{\odot}\, yr^{-1}$.

If, as before, the old stellar population has a significant effect on the dust heating at these wavelengths, we would naively expect to see a correlation between the FIR luminosity and 3.6\,\micron\, emission. However, the total FIR luminosity is a function of dust mass and dust temperature, so if the distribution of dust is different to the distribution of stars (as it is in M31, \citealt{smith2012b}) there will be no correlation, even if the old stars are the major heating source \citep{bendo2012}. This was tested and no correlation is visible.

Without any kind of correction for the old stars, the star formation rate from the FIR emission is measured to be approximately double the estimate from the FUV and 24\,\micron\, tracers (Section \ref{sec:fuv24}). This is expected as M31 has not gone through a starburst in its recent history, so a significant portion of the heating is due to the interstellar radiation field (ISRF).

As discussed in the previous section, past work on M31 elected to omit the central region of the galaxy when determining the global SFR, due to the dominance of old stars in this region. If we omit the central region out to $0.2\,{\rm R_{M31}}$ the measured SFR reduces from $0.52\,\rm M_{\odot}\,yr^{-1}$ to $0.48\,\rm M_{\odot}\,\rm yr^{-1}$. This minimal difference suggests that the over-estimate is not limited to heating from old stars in the bulge. This is consistent with the correlation observed between 3.6\,\micron\, emission and 24\,\micron\, in the ring (Section \ref{sec:fuv24}), indicating old stars have a significant heating effect here also.

\subsection{Comparison of star formation tracers}
\label{sec:comp}
\begin{figure*}[!htbp]
  \centering
    \includegraphics[width=0.55\textwidth, trim=0 180 1151.1 1378, clip=true]{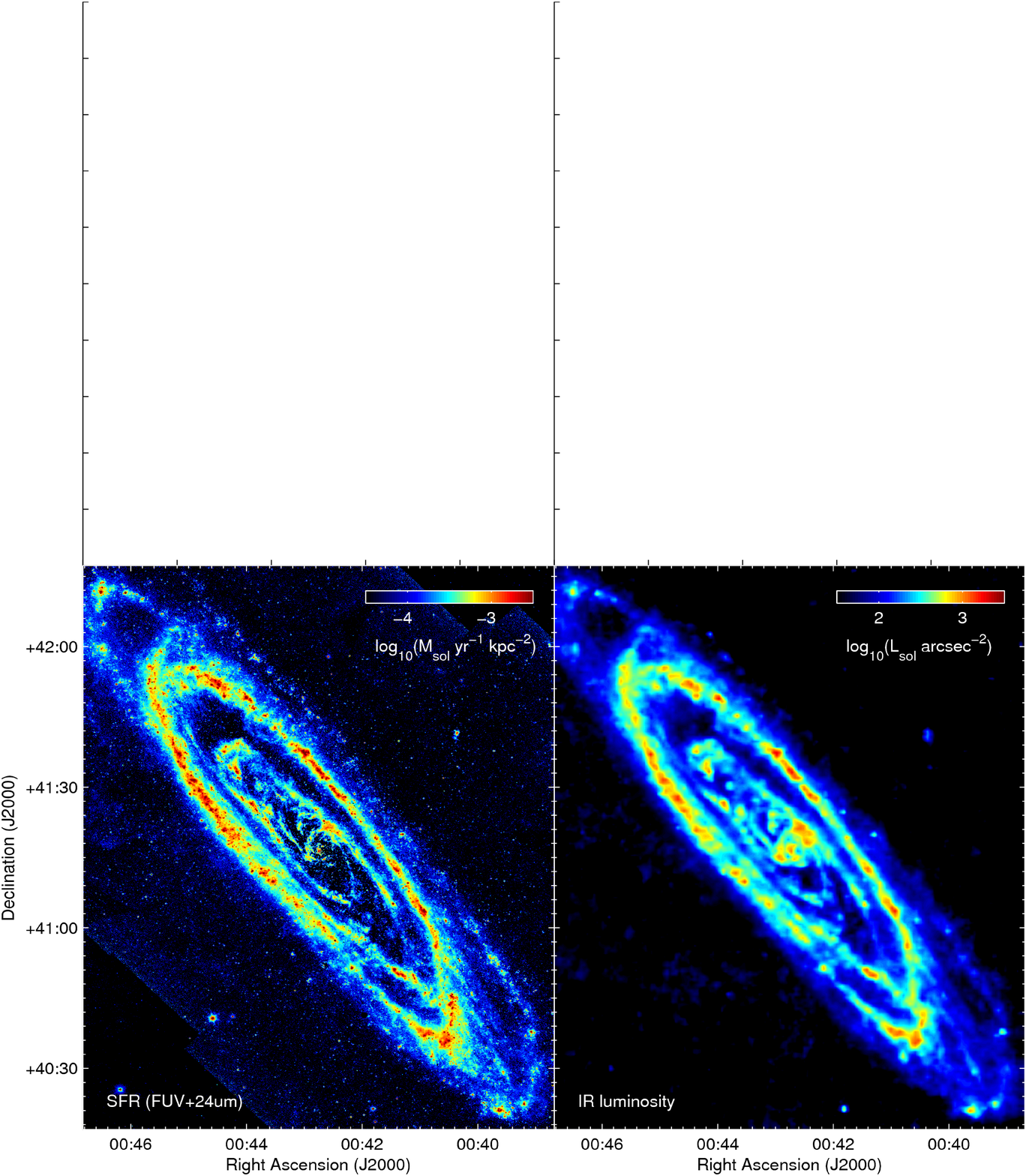}\includegraphics[width=0.45\textwidth, trim=1406.9 180 0 1378, clip=true]{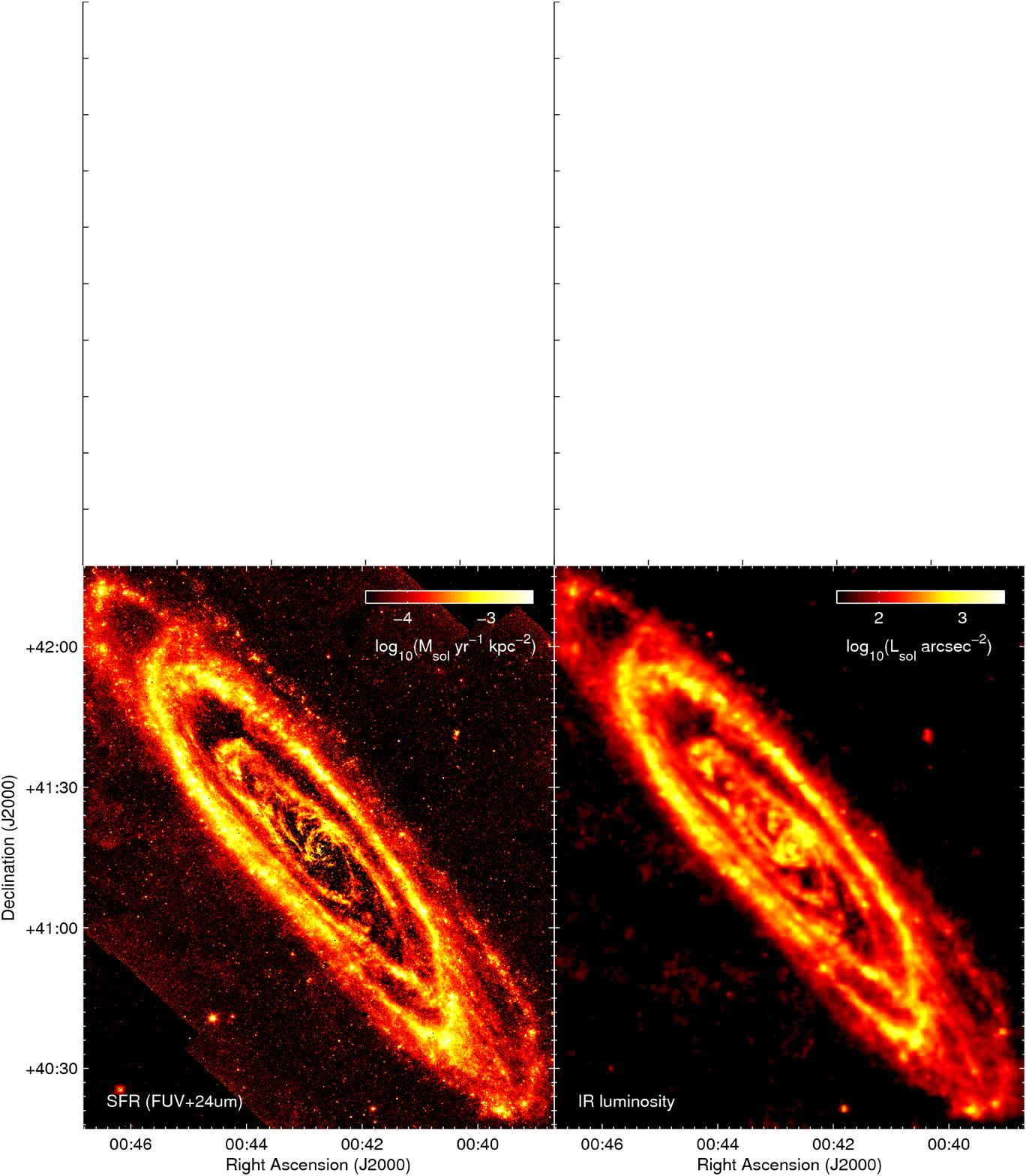}
  \caption{Left, star formation rate map obtained using the FUV and 24\,\micron\, emission ({\sc FWHM} beamwidth = 6\,\arcsec, pixel size = 1.5\,\arcsec); right, total FIR luminosity at $\lambda>70\,\micron$ ({\sc FWHM} beamwidth and pixel size = 36\,\arcsec), found by integrating the emission from {\it Herschel} and {\it Spitzer} observations shown in Figure \ref{fig:mapsIR}. The SFR from infrared luminosity is found using a constant conversion factor based on the assumed IMF and assumed length of the continuous starburst.}
\label{fig:mapssfr}
\end{figure*}

The star formation maps made using FUV and 24\,\micron\, emission, and that from FIR luminosity can be seen in Figure \ref{fig:mapssfr}. The FUV and 24\,\micron\, tracer has units of $\rm M_{\odot}\,yr^{-1}\,\rm kpc^{-2}$, but we have elected to display the FIR star formation tracer in terms of FIR luminosity as the conversion factor between luminosity and star formation rate is very uncertain.

% FIGURE 5

Sub-mm wavelengths are more greatly affected by heating due to the ISRF than the 24\,\micron\, emission, as this regime is sensitive to cooler dust temperatures. This means the FIR emission is susceptible to heating from more distant stars (those that are not in the same pixel), making determination of a correction factor difficult (Section \ref{sec:sfr_ir}). In order to compare the two SFR tracers, we modify the FUV and 24\,\micron\, tracer to match the assumptions made in creating the map of $\Sigma_{\rm SFR}$ from FIR luminosity.

\begin{figure*}[!htbp]
  \centering
    \includegraphics[width=\textwidth]{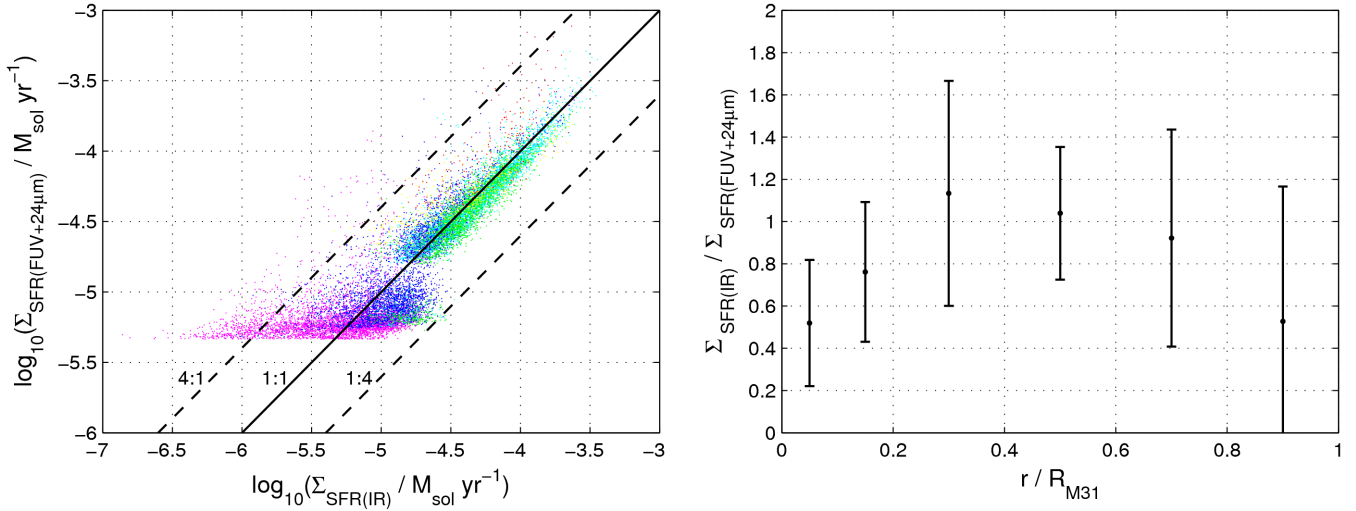}
  \caption[Comparison of star formation tracers]{Left: $\Sigma_{\rm SFR}$ (star formation rate surface density) found from FIR luminosity (assuming a Salpeter IMF) vs $\Sigma_{\rm SFR}$ from FUV and 24\,\micron\, emission (scaled to match values assuming a Salpeter IMF and without a correction for the old stars). The colours represent the radius of the pixel from the galactic centre (Figure \ref{fig:mapsgas}). The solid black line indicates a 1:1 relationship, the dashed black lines indicate factors of 4 offsets. Right: ratio of $\Sigma_{\rm SFR}$ from FIR luminosity to $\Sigma_{\rm SFR}$ from FUV+24\,\micron\, star formation surface density with radius. The errorbars represent one standard deviation of the scatter in this value across each elliptical annulus.}
\label{fig:IR_v_Leroy}
\end{figure*}

\citet{leroy2008} state that their tracer returns a star formation rate $1.59\times$ lower than \citet{kennicutt1998b}, from where we get our $\delta_{\rm MF}$ (Equation \ref{eq:sfr_ir}). Because of this, in Figure \ref{fig:IR_v_Leroy} we compare $\Sigma_{\rm SFR}$ from FIR luminosity (as described in Section \ref{sec:sfr_ir}) with the same from FUV and 24\,\micron\, with an additional factor of 1.59 to approximate the same IMF, and no correction for old stars. The low-mass cut-off ($m_{\rm 1} = 0.1\,\rm M_{\odot}$) of the IMF and star formation timescale ($\tau = 100\,\rm Myr$) are already equivalent. This gives a global SFR of $0.51\,\rm M_{\odot}\,yr^{-1}$, consistent with the value from FIR luminosity.

% FIGURE 6

It does appear that the SFR as measured from FIR luminosity is slightly lower relative to the FUV and 24\,\micron\, tracer in the very centre and outer regions of M31 (Figure \ref{fig:IR_v_Leroy}, right). There is a possible issue with \emph{PACS} observations not recovering all of the flux in low surface brightness regions \citep[e.g.][]{aniano2012}. If this was the case, it may contribute to the observed effect between annuli but our global SFR will be minimally affected. We do not believe this discrepancy is a major issue here as this map is not used in any further resolved analyses of low surface brightness regions.

Despite the general consistency between SFR tracers, we should remain aware that since the conversion factor between tracer luminosity and star formation rate depends on the IMF, we are not necessarily recovering the correct value. It is possible that the IMF we assumed for M31 is not appropriate, or, because the star formation rate is low, we are not sampling the whole IMF leading to fluctuations in the tracer luminosity for a fixed SFR. This can be an issue for a variety of tracers, including IR and UV emission \citep{kennicutt2012}.

For the analysis that follows, we elected to use the combined FUV and 24\,\micron\, emission as our star formation tracer, as we are able to correct for the old stellar population. We argue that this gleans more reliable SFRs in low SFR regimes, like those in M31, than when using the FIR luminosity.

\section{The interstellar medium in M31}
\label{sec:gas}
The ISM is made of predominantly neutral atomic and molecular hydrogen. A map of total gas can be produced by summing these two constituents and multiplying by a factor of 1.36 to account for heavier element abundances (mostly Helium), or alternatively by assuming the total gas is well traced by dust emission \citep[e.g.][]{eales2012}. 

\subsection{Total gas from H\,{\sc i} and CO observations}
The H\,{\sc i} map is taken from \cite{braun2009}. In order to keep consistency with our maps of star formation, and to allow comparison of galaxies with different inclinations, $i$, we employ a factor of $\cos{i}\,$ to `deproject' the galaxy.

\label{sec:COtoH2}
H$_{2}$ is the most abundant molecule in the ISM, but lacks a dipole moment so is not easily observable. For this reason, CO (usually $J$=1-0) is employed as a tracer molecule. For M31 we use the map from \citet{nieten2006}. The conversion between CO emission and quantity of H$_{2}$ is still a contentious topic and uses the so-called $X$-factor \citep[e.g.][]{wall2007, glover2011, narayanan2011, feldmann2011, bolatto2013}, where:
\begin{equation}
N_{\rm H_{2}} / {\rm cm^{-2}} = X \times I_{\rm CO} / {\rm K}\,{\rm km}\,s^{-1}
\end{equation}

The conversion factor specific to molecular clouds in the north eastern arm of M31 was argued to be $5.68 \times 10^{20} \,(\rm K\,km\,s^{-1})^{-1}\,cm^{-2}\,$ in \citet{sofue1994a}. This was found by estimating virial masses (from their size and velocity width) and comparing to the CO line intensity. This is larger than the value found by \citet{bolatto2008} of $\sim4\times 10^{20} \,(\rm K\,km\,s^{-1})^{-1}\,cm^{-2}\,$ using the same method. However, the ISM of M31 is dominated by neutral atomic hydrogen so it is not clear whether the virial masses provide an overestimate of the mass of molecular hydrogen in these clouds. Here we will assume $X=2\times10^{-20} \,(\rm K\,km\,s^{-1})^{-1}\,cm^{-2}$ \citep[e.g.][]{strong1988, pineda2010}, which agrees with the value derived in \citet{smith2012b}. Any constant discrepancy in the $X$-factor will result in a horizontal translation in our $\rm log_{10}(\Sigma_{H_{2}})$ vs $\rm log_{10}(\Sigma_{SFR})$ plots and so will have no effect on the calculation of our K-S index for molecular gas. It may skew the calculation using total gas but the effect is likely to be small due to the dominance of H\,{\sc i} in the ISM of M31.

At this point we should note the suggestion that metallicity has an effect on the $X$-factor \citep[e.g.][]{israel1997, strong2004}. \citet{smith2012b} found a radial variation in the gas-to-dust ratio suggesting a metallicity gradient in M31 and hence a gradient in $X$. Any variation should not be a big issue when looking at total gas, as the ISM in M31 is dominated by neutral atomic hydrogen. We do need to keep this uncertainty in mind when dealing with molecular gas, however.

\subsection{Total gas traced by dust}
\label{sec:dust}
The interstellar gas can, in principle, also be traced by the distribution of dust in a galaxy \citep{eales2012}. The dust map of M31 is taken from \citet{smith2012b}, where dust mass is found by fitting a modified blackbody function in each pixel where there is a $5\,\sigma\,$ detection in all bands. This should mitigate against the low-surface brightness issues discussed in Section \ref{sec:comp}.

In \citet{smith2012b} a fit was performed to gas-to-dust vs radius, to determine how the conversion factor varies. It was found that the relationship is logarithmic with radius varying between $\sim30$ near the centre and $\sim100$ in the 10$\,{\rm kpc}$ ring, consistent with the value found in the MW \citep{spitzer1978}. We use this function to create a second total gas map as traced by dust. In the following section, as with the other ISM tracers, this will be used to observe how well dust mass correlates with star formation.

\newpage
\section{The star formation law}

In this section, we probe the star formation or K-S law assuming the following relationship, 
\begin{equation}
\Sigma_{\rm SFR} = A\,\Sigma_{\rm Gas}^{N},
\label{eq:schmidt}
\end{equation}
where $N$ is the power index and $A$ is related to the star formation efficiency (SFE).

We separately look at how M31 compares to other local galaxies in terms of global SFR surface density (calculated from FUV and 24\,\micron\, emission) and gas surface density; and what relationship the star formation law follows on a pixel-by-pixel basis when considering various components of interstellar material.

\subsection{Global star formation law}

\begin{figure*}[!htb]
  \centering
    \includegraphics[width=\textwidth]{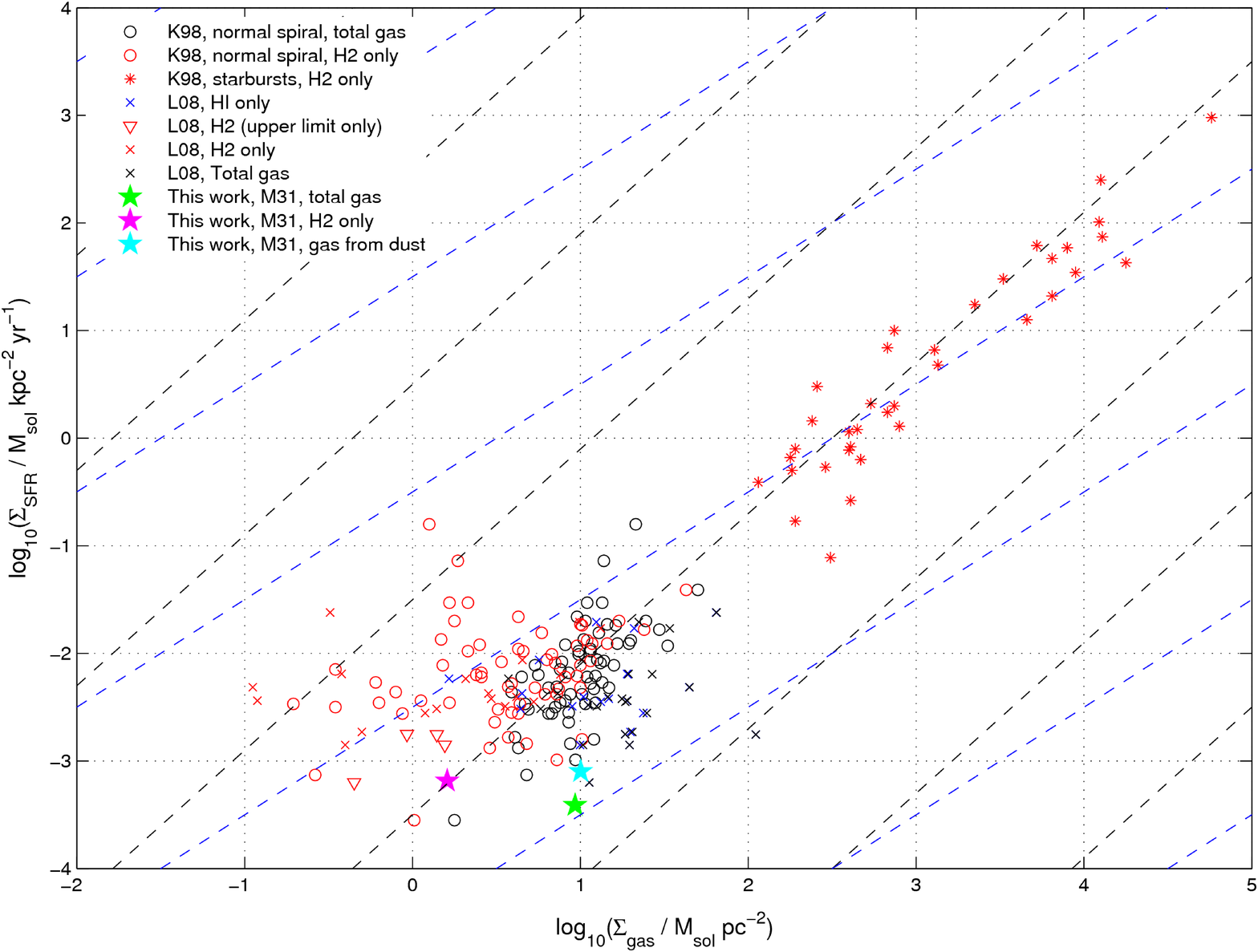}
  \caption[Comparison with previous work]{Global SFR vs gas mass, derived in M31 using three gas tracers, with \citet{kennicutt1998b} and \citet{leroy2008} galaxies. Galaxies are plotted using total gas, H\,{\sc i} only and H$_{2}$ only, depending on availability of data. SFRs from this work and \citet{leroy2008} are scaled to match the assumptions of \citet{kennicutt1998b}. The dashed diagonal lines represent the gradient the galaxies should follow given a Schmidt law of the type found in \citet{kennicutt1998b} (N = 1.4, black) or a linear relationship (N = 1, blue), as has been suggested by several recent papers \citep[e.g.][]{rahman2012}.}
\label{fig:ford_kennicutt_leroy}
\end{figure*}

Figure \ref{fig:ford_kennicutt_leroy} compares the mean surface density of star formation rate with the mean surface density of gas for global measurements of galaxies from \citet{kennicutt1998b} and \citet{leroy2008}, with corresponding global values for M31 overplotted. The SFRs from this paper and \citet{leroy2008} are scaled to match the assumptions made in \citet{kennicutt1998b}. The mean values for M31 are found over all pixels with sufficient signal-to-noise in both maps (SFR and gas). The difference in measured SFR is due to the different selection effects depending on the gas tracer. We can immediately see that the mean SFR for regions containing sufficient H$_{2}$ ($I_{\rm CO} > 5\,\sigma_{\rm CO}$) or dust ($I > 5\,\sigma$ in five \emph{Herschel} bands) is higher, suggesting a better spatial correlation between SFR and both molecular hydrogen and dust than total gas.

% FIGURE 7

The low gas mass galaxies studied by \citet{kennicutt1998b} generally appear to have higher star formation rates than M31, although we note that they estimate star formation using a different SF tracer. However, early-type spirals like M31 are expected to exhibit a low SFR per unit area, as stated in \citet{kennicutt1998a}. Mean surface densities of both total and molecular gas are consistent with the same parameters for normal spirals studied in previous work.

\subsection{Resolved star formation law}
\label{sec:schmidt}
From our gas and SFR maps, we should be able to investigate the Kennicutt-Schmidt star formation law, on a `per pixel' basis across the galaxy. This section aims to test how calculation of this law changes with different gas tracers. It has been suggested that H$_{2}$ is a better tracer of star formation than total gas \citep[e.g.][]{bigiel2011}, although it is not clear if this is the case in M31 where H\,{\sc i} dominates the ISM.

Here we use data from our maps of surface density of star formation as found from FUV and 24\,\micron\, emission, against surface density of total gas, molecular hydrogen only and gas traced by dust. We select pixels that satisfy $\Sigma_{\rm SFR} > 5\,\sigma_{\rm SFR}\,$ and $\Sigma_{\rm Gas} > 5\,\sigma_{\rm Gas}$, where $\sigma_{\rm SFR}$ is the standard deviation of the background of the star formation map and $\sigma_{\rm Gas}$ is a combination of the uncertainties of the constituent gas maps (e.g. for total gas this will be the scaled uncertainties in the integrated H\,{\sc i} and CO($J$=1-0) images).

% FIGURE 8
\begin{figure*}[!htb]
  \centering
    \includegraphics[width=\textwidth, trim=30 20 0 0, clip=false]{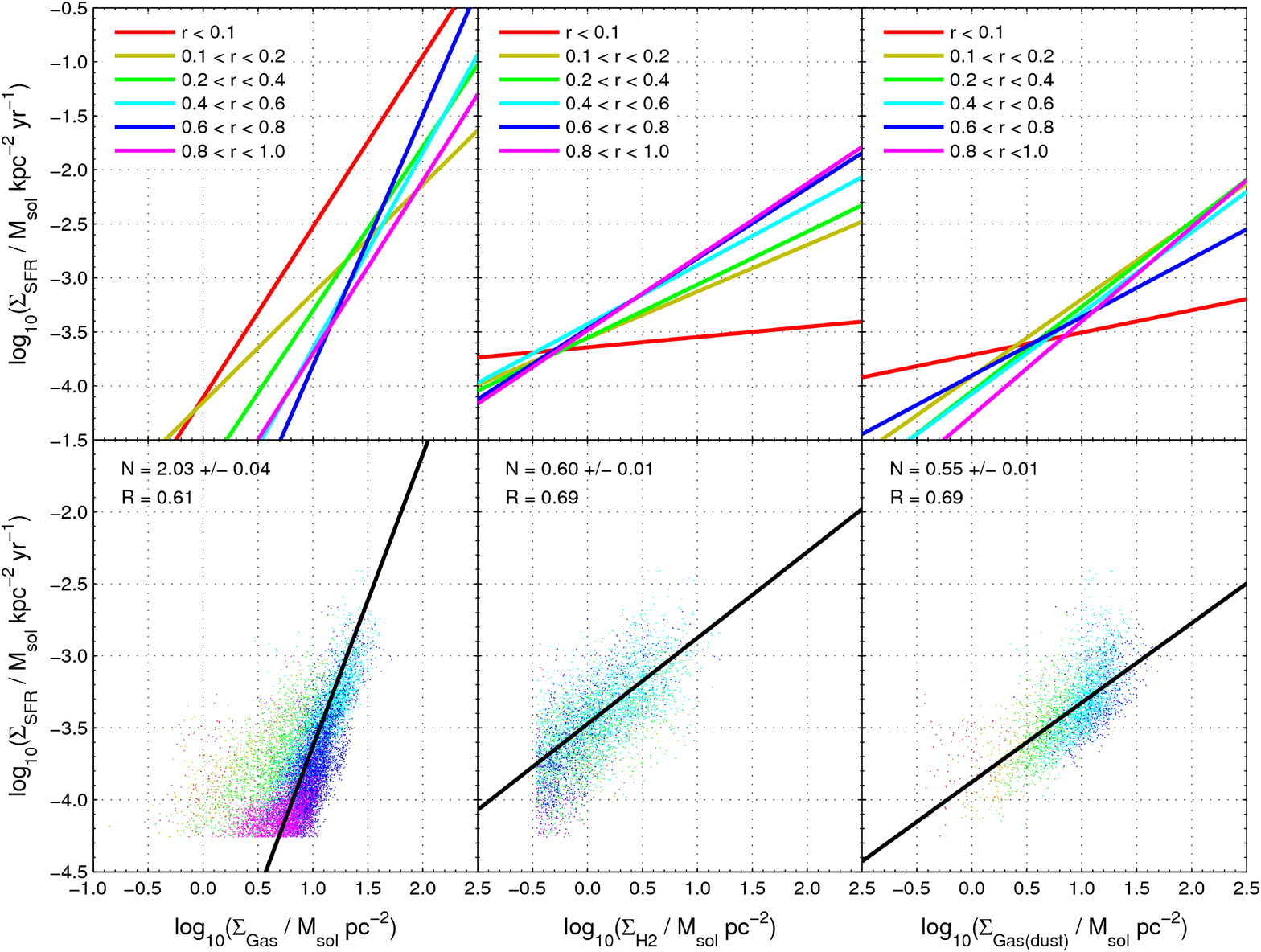}
  \caption[SFR vs Gas in M31]{Star formation rate surface density against gas surface density for three gas tracers. The top row shows fits to the Kennicutt-Schmidt law for each annulus. The bottom row shows the data for every pixel studied, with a global fit indicated by a solid black trendline, with the calculated index $N$ quoted in the top left corner of each panel along with the correlation coefficient of unbinned data, $R$. Each column denotes a single gas tracer. From left, they are: total gas from H\,{\sc i} and CO($J$=1-0), assuming a CO-H$_{2}$ conversion factor of $2\times10^{-20} \,(\rm K\,km\,s^{-1})^{-1}\,cm^{-2}$, with an additional factor of 1.36 for heavier element abundances; molecular hydrogen (H$_{2}$) from CO($J$=1-0); and total gas traced by dust mass \citep[see][]{smith2012b}, assuming a radial gradient in the gas-to-dust ratio.}
\label{fig:schmidt}
\end{figure*}

\subsubsection{Fitting}
Here we perform a linear fit in order to find the index, $N$ from equation \ref{eq:schmidt}, assuming that
\begin{equation}
{\rm log_{10}}\,\Sigma_{\rm SFR}\, \propto N\,{\rm log_{10}}\,\Sigma_{\rm Gas}.
\label{eq:schmidt_log}
\end{equation}

In Figure \ref{fig:schmidt}, the signal-to-noise (S/N) cuts are clearly manifest. In total gas, the major cut-off is horizontal (limited by $\sigma_{\rm SFR}$); in molecular gas the cut is vertical (limited by $\sigma_{\rm H_{2}}$). Previous works appear to exhibit a similar cut-off \citep[e.g.][]{tabatabaei2010} but with no attempt to mitigate for this when performing a fit. After some exploration we conclude that the signal to noise cut does indeed bias the data and must be mitigated against (see Appendix). We test other methods of fitting and find that binning the data in order of increasing star formation gives the most reliable return gradient when testing the relationship to total gas.

We therefore attempt to mitigate for the S/N cut in our data using the same method. When looking at the total gas from H\,{\sc i} and CO measurements, we order in bins of increasing SFR, with an equal number of datapoints (500) in each. We then plot the mean surface density of gas ($\Sigma_{\rm Gas}$ / $\rm M_{\odot}\,kpc^{-2}$) in each bin, against the mean surface density of SFR ($\Sigma_{\rm SFR}$ / $\rm M_{\odot}\,kpc^{-2}\,yr^{-1}$) and perform the fit on these points in the logarithmic domain using a least squares routine in \emph{MATLAB}.

In the case of H$_{2}$ only, the S/N cut-off is more apparent in gas mass so we bin the data in order of increasing gas mass, with 100 points in each bin.

Gas mass estimated from dust mass exhibits a more complex selection effect so binning is not attempted here. The majority of the points that are omitted correspond to points that appear towards the low-SFR regime of the bottom-left window in Figure \ref{fig:schmidt}. This will affect our calculation of the SF law but we believe the analysis is still valid as we are preferentially selecting regions that are more important for star formation.

\subsubsection{Kennicutt-Schmidt index}
\begin{figure*}[!htb]
  \centering
    \includegraphics[width=\textwidth, trim=100 0 100 0, clip=true]{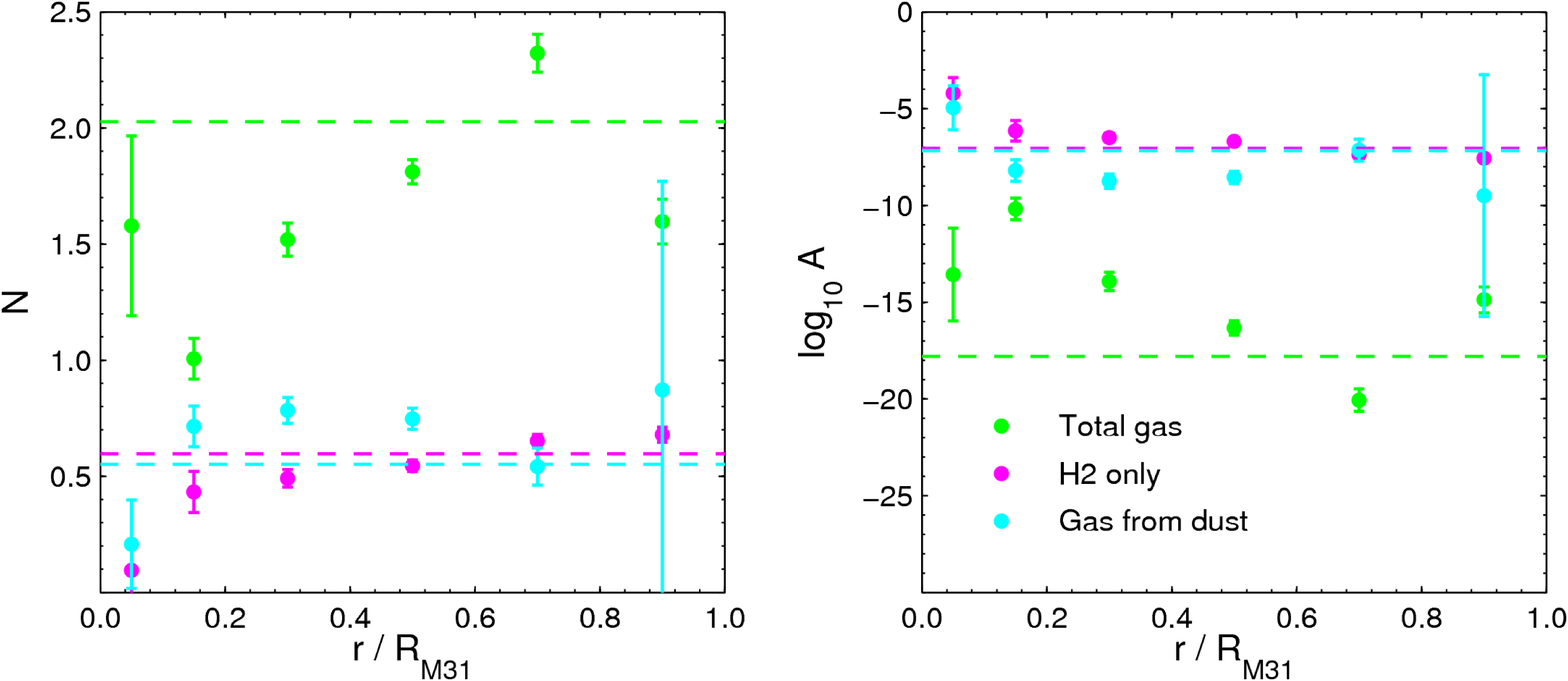}
  \caption[Kennicutt-Schmidt parameters with radius]{Kennicutt-Schmidt parameters with radius across M31. We compare the power law indices, $N$ and parameter $A$ using H\,{\sc i} + H$_{2}$, H$_{2}$ only and total gas mass traced by dust. The dashed lines indicate the global values for M31. Errorbars indicate the $2\sigma$ uncertainty based on the binned fitting.}
\label{fig:schmidt_v_r}
\end{figure*}

Figure \ref{fig:schmidt} shows plots of SFR versus gas mass (in units of $\rm M_{\odot}\,pc^{-2}$, to keep consistency with previous work) for each pixel, with trendlines for each annulus (top row) and a global fit (bottom row). Figure \ref{fig:schmidt_v_r} shows K-S index as a function of radius for the various gas tracers. Global values are indicated by a dashed line and are also given in Table \ref{tab:schmidt}.

% FIGURE 9

\begin{deluxetable}{l r r r r r }
  \tabletypesize{\scriptsize}
  \tablecaption{Global K-S parameters using different tracers.}
  \tablewidth{0.50\textwidth}
  \tablehead{\colhead{Gas tracer} & \colhead{$N$} & \colhead{$\sigma_{N}$} & \colhead{log(A)} & \colhead{$\sigma_{A} / {\rm dex}$} & \colhead{$R$}}
  \startdata
    Total gas (H\,{\sc i} + CO) & 2.03 & 0.04 & -19.11 & 0.31 & 0.6\\
    H$_{2}$ only from CO & 0.60 & 0.01 & -7.41 & 0.08 & 0.7\\
    Total gas from dust & 0.55 & 0.01 & -7.55 & 0.07 & 0.7
  \enddata
  \tablecomments{$\Sigma_{\rm SFR} = A\,\Sigma_{\rm Gas}^{N}$. The quoted standard deviation refers to that found using the fit on binned data. The correlation coefficient, $R$ refers to raw data points.}
\label{tab:schmidt}
\end{deluxetable}

The K-S index for each annulus varies between 1.0 and 2.3 when considering total gas, with the higher values applying to the 10$\,{\rm kpc}$ ring. The global value for total gas is $\sim2.0$, decreasing to $\sim1.8$ if we double the $X$-factor.

The star formation law with H$_{2}$ gives a shallower gradient ($N = 0.6$) but is more constant between annuli. Doing the same with gas traced by dust gives a similarly shallow slope.

\subsection{Discussion}
Our star formation law in M31 using total gas gives a K-S index $N\sim2.0$, significantly higher than the value found by \citet{tabatabaei2010} but consistent with values found in previous work on other galaxies. One possible explanation for a steep slope is H\,{\sc i} becoming optically thick. However, when performing the analysis on an opacity corrected map of H\,{\sc i}, we see no real change in the calculated star formation law. An alternative is that the hydrogen turns molecular (which occurs at $\Sigma_{\rm Gas} \sim 10\,\rm M_{\odot}\,pc^{-2}$) but is not traced by our CO($J$=1-0) map in these high SFR regimes, hence the turnover apparent in previous work \citep[e.g.][]{bigiel2008} is not visible here.

We get a lower mean star formation rate when looking at all regions of gas which suggests we are not isolating star forming regions as well as when using other gas tracers. The positions of datapoints for each pixel in Figure \ref{fig:schmidt} (bottom row) for total gas, appear to depend on their radius. This is consistent with the fits to each annulus showing a horizontal offset (top row), with the inner regions to the left of the plot (low gas mass). This would suggest that the threshold for star formation changes with radius (see also Figure \ref{fig:schmidt_v_r}). However we should note that the inner regions contain relatively fewer datapoints. Also, it is possible that despite our correction, we still overestimate star formation in the centre.

The index found here using molecular gas ($N \sim 0.6$) does argue against a superlinear relationship on small scales. It is possible the resolution of our images may affect calculation of our index but when testing the relationship using a 500 pc pixel scale, we see a negligible difference. Some recent work has suggested that the star formation law is linear ($N = 1$) when looking at molecular clouds \citep[e.g.][]{rahman2012} and that the superlinear relationship of \citet{kennicutt1998b} is not a manifestation of a relationship that applies on smaller scales, but may be the result of systematic differences between the galaxies that are related, but not limited to, gas mass alone.

A `sub-linear' relationship ($N < 1$), indicates that star formation is less efficient at high gas densities which would be an intriguing result. We urge caution with our value for the index with molecular gas however, due to the significant scatter.

One other issue we should keep in mind is that because the scales we are probing are small compared to other extragalactic sources, the surface density of star formation and surface density of gas may not be directly relatable to the corresponding volume densities as our scale height is more likely to vary between regions.

\section{Summary}
In this paper we have determined the surface density of star formation in M31 using combined FUV and 24\,\micron\, emission and separately the far-infrared luminosity. We aim to correct the former for emission from both unobscured and embedded old stars and find a global star formation rate of $0.25^{+0.06}_{-0.04}\,\rm M_{\odot}\,yr^{-1}$. The FIR emission appears to be correlated with the SFR map made using FUV and 24\,\micron\, emission. However, we are unable to correct for the old stellar population as there is no correlation visible between FIR luminosity and 3.6\,\micron\, emission in the galactic centre.

We produce two maps of the total gas in M31. The first uses H\,{\sc i} and CO, assuming a CO-H$_{2}$ conversion factor of $2\times10^{20} \,(\rm K\,km\,s^{-1})^{-1}\,cm^{-2}$. We use the radially varying gas-to-dust ratio found in \citet{smith2012b} to produce the second map of total gas from the dust emission.

When comparing with previous work by \citet{kennicutt1998b} and \citet{leroy2008} on the global SFR and gas mass, we find the mean molecular gas surface density and star formation rate surface density for M31 sit on the low end of the relation determined in \citet{kennicutt1998b}.

Our measurement of the star formation law on sub-kpc scales varies with gas tracer. The most direct measurement, using H\,{\sc i} and CO to trace total gas, gives power law index  $N \sim 2.0$ when looking at the whole galaxy, consistent with the range of values found in previous work but we believe this slope is a result of H\,{\sc i} saturation. The values measured in radial annuli vary between 1.0 and 2.3, with the highest values being measured in the 10 kpc ring, where the vast majority of star formation is occurring.

Using molecular gas only gives a much lower K-S index of $N \sim 0.6$, suggesting that a superlinear relationship with molecular gas is not applicable on sub-kpc scales in M31.

\section*{Acknowledgments}
Thanks to Pauline Barmby for use of her 3.6\,\micron\, map of M31 and to Luca Cortese for his advice regarding GALEX data.

We thank everyone involved with the Herschel Observatory. PACS has been developed by a consortium of institutes led by MPE (Germany) and including UVIE (Austria); KU Leuven, CSL, IMEC (Belgium); CEA, LAM (France); MPIA (Germany); INAF- IFSI/OAA/OAP/OAT, LENS, SISSA (Italy); and IAC (Spain). This development has been supported by the funding agencies BMVIT (Austria), ESA-PRODEX (Belgium), CEA/CNES (France), DLR (Germany), ASI/INAF (Italy), and CICYT/MCYT (Spain).
   SPIRE has been developed by a consortium of institutes led by Cardiff University (UK) and including: the University of Lethbridge (Canada); NAOC (China); CEA, LAM (France); IFSI, University of Padua (Italy); IAC (Spain); Stockholm Observatory (Sweden); Imperial College London, RAL, UCL-MSSL, UKATC, University of Sussex (UK); and Caltech, JPL, NHSC, and the University of Colorado (USA). This development has been supported by national funding agencies: CSA (Canada); NAOC (China); CEA, CNES, CNRS (France); ASI (Italy); MCINN (Spain); SNSB (Sweden); STFC, UKSA (UK); and NASA (USA).

We finally thank the anonymous referee for their helpful suggestions that have undoubtedly improved our paper.

\newpage
%\bibliographystyle{apj}
%{plain}
%\bibliography{papers}

\appendix
\section{Fitting}
\label{sec:fitting}
Our $\Sigma_{\rm Gas}$ against $\Sigma_{\rm SFR}$ plots exhibit clear signal-to-noise (S/N) cut offs which it is imperative we address. The first task was to test whether the cuts result in a bias when performing the fit.

\begin{figure}[!htb]
  \centering
    \includegraphics[width=\textwidth, trim=150 0 200 50, clip=true]{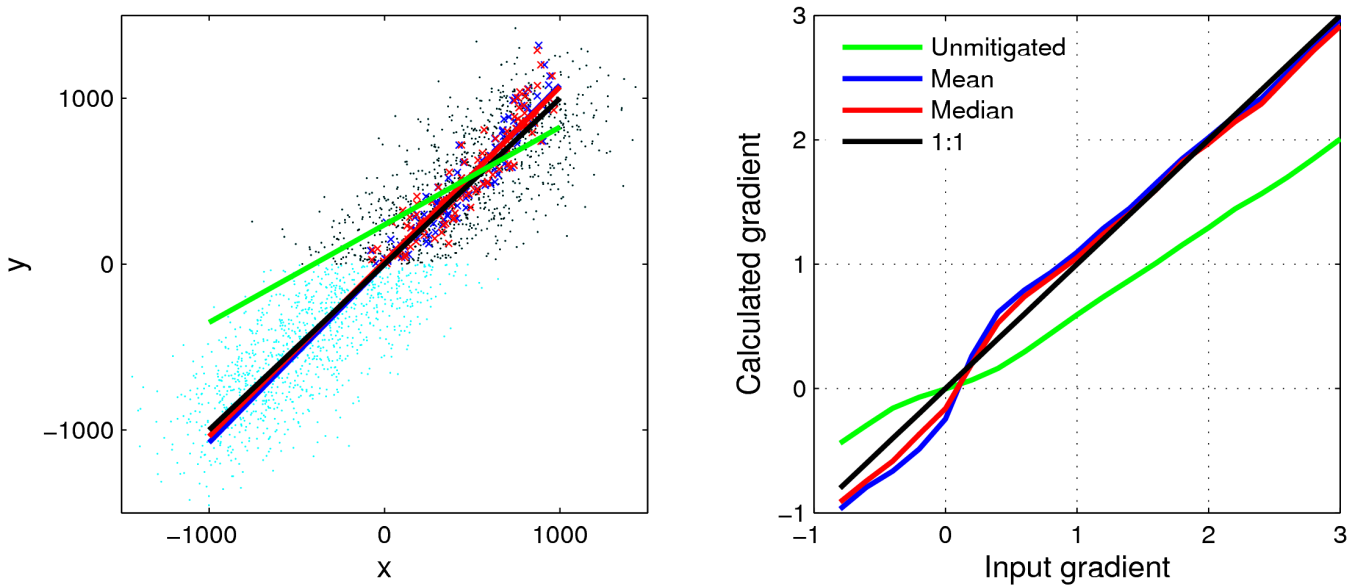}
  \caption[K-S index fitting]{Simulated Kennicutt-Schmidt index fitting. Left: Example simulated dataset with input gradient of 1.0. Black points are the selected data, cyan points are those that have been discarded before performing the fit. Right: Input gradient versus measured gradient for a range of input gradients and fitting methods. In both plots, green represents the unmitigated fit; blue represents the mean of binned data; red, the median. The black trendlines represent the input gradient.}
\label{fig:fitting}
\end{figure}

We created two arrays, $x$ and $y$ where $x$ contains all integers between $-1000$ and $+1000$ and $y = mx$ where $-1 < m < +3$. Gaussian noise is applied to both the $x$ and $y$ values to simulate the observed spread in points. We then apply a cut at a specified $y$ value, again mimicking the data (Figure \ref{fig:fitting}).

The \emph{polyfit} algorithm in \emph{MATLAB} is used to perform the fit on the data above the cut. \emph{Polyfit} is a least-squares routine that minimises residuals in the y-axis parameter. When $m \neq 0$ the calculated gradient is consistently shallower than the input, indicating a bias. We attempt to mitigate for this by ordering the data in bins of increasing $y$ with an equal number of points in each bin. We replace this data with a single point based on the the mean or median of the binned data. The fit is then performed using the same algorithm on these averaged points. In both cases we get slightly steeper gradients when $-1 \lesssim m \lesssim +1$ converging to near perfect agreement in the region where $1.5 \lesssim m \lesssim 2.0$. In all cases studied, the unmitigated fit is more deviant.

We therefore attempt to mitigate for the S/N cut in our data using the same method. When looking at the total gas from H{\sc i} and CO measurements, we order in bins of increasing SFR, with an equal number of datapoints (500) in each. We then plot the mean gas mass in each bin, against the mean SFR and perform the fit on these points, using the logarithmic units.

In the case of H$_{2}$ only, the SNR cut-off is more apparent in gas mass so we bin the data in order of increasing gas mass, with 100 points in each bin. Gas mass as estimated from dust mass exhibits a more complex selection effect so binning is not attempted here.

\end{document}